\documentclass[aip,apl
sd,amsmath,amssymb,reprint]{revtex4-1}

\usepackage{graphicx}
\usepackage{dcolumn}
\usepackage{bm}

\begin{document}

\preprint{}

\title[Adiyatullin \textit{et. al.}]{Temporally resolved second-order photon correlations of exciton-polariton Bose-Einstein condensate formation}

\author{Albert F. Adiyatullin}
	\email{albert.adiyatullin@epfl.ch}
	\affiliation{Laboratory of Quantum Optoelectronics, \'{E}cole Polytechnique F\'{e}d\'{e}rale de Lausanne, CH-1015, Lausanne, Switzerland}
\author{Mitchell D. Anderson}
	\affiliation{Laboratory of Quantum Optoelectronics, \'{E}cole Polytechnique F\'{e}d\'{e}rale de Lausanne, CH-1015, Lausanne, Switzerland}
\author{Pierre V. Busi}
	\affiliation{Laboratory of Quantum Optoelectronics, \'{E}cole Polytechnique F\'{e}d\'{e}rale de Lausanne, CH-1015, Lausanne, Switzerland}
\author{Hadis Abbaspour}
	\altaffiliation[Present address: ]{Institute of Quantum Electronics, ETH Zurich, CH-8093 Zurich, Switzerland}
	\affiliation{Laboratory of Quantum Optoelectronics, \'{E}cole Polytechnique F\'{e}d\'{e}rale de Lausanne, CH-1015, Lausanne, Switzerland}
\author{R\'egis Andr\'e}
	\affiliation{Institut N\'eel, CNRS, 25 Avenue des Martyrs, 38042 Grenoble, France}
\author{Marcia T. Portella-Oberli}
	\affiliation{Laboratory of Quantum Optoelectronics, \'{E}cole Polytechnique F\'{e}d\'{e}rale de Lausanne, CH-1015, Lausanne, Switzerland}
\author{Benoit Deveaud}
	\affiliation{Laboratory of Quantum Optoelectronics, \'{E}cole Polytechnique F\'{e}d\'{e}rale de Lausanne, CH-1015, Lausanne, Switzerland}

\date{\today}

\begin{abstract}
Second-order time correlation measurements with a temporal resolution better than 3 ps were performed on a CdTe microcavity where spontaneous Bose-Einstein condensation is observed. After the laser pulse, the nonresonantly excited thermal polariton population relaxes into a coherent polariton condensate. Photon statistics of the light emitted by the microcavity evidences a clear phase transition from the thermal state to a coherent state, which occurs within 3.2 ps after the onset of stimulated scattering. Following this very fast transition, we show that the emission possesses a very high coherence that persists for more than 100 ps after the build-up of the condensate.
\end{abstract}

\maketitle

Exciton-polariton Bose-Einstein condensates (BEC) are a prime example of a macroscopic quantum state which emit coherent light over a large spatial region \cite{Kasprzak_Nature2006,Snoke_Sci2002,Richard_PRL2005,Balili_Science2007,Nardin_PRL2009,Whittaker_EUROLETT2009,Deng_RMP2010,Keeling_CP2011}. Particularly, polariton BECs can be formed after nonresonant excitation above the threshold power with spontaneous formation of a macroscopic coherence in the lowest energy state \cite{Kasprzak_Nature2006}. This has been evidenced using first-order spatial correlation \cite{Kasprzak_Nature2006} and time-correlation measurements\cite{Love_PRL2008}, as well as through the study of the second-order time correlation function ($g^{(2)}$) with Hanbury Brown and Twiss (HBT) detection scheme\cite{Kasprzak_PRL2008,Love_PRL2008}. Later, the dynamics of BEC formation using the time-resolved optical interferometry was shown \cite{Nardin_PRL2009}. However, access to time resolved $g^{(2)}$ during the relaxation of the polaritons towards the BEC was not possible because it has a typical timescale of only a few ps, and even the best present semiconductor detectors do not have sufficient resolution \cite{APD_review}.

Recently, a new technique for performing $g^{(2)}$ measurements was developed \cite{Wiersig_Nat2009,Assmann_OE2010,Takemura_PRA2012,Silva_arXiv2014}, taking advantage of a streak-camera as a photodetector giving temporal resolution on the picosecond timescale. This approach has already been used to demonstrate the difference between the thermal state of polaritons and a polariton BEC \cite{Tempel_PRB2012} as well as between the regimes of strong and weak coupling \cite{Assmann_sci2009,Tempel_PRB2012}. However, the dynamics of the relaxation of thermal polaritons into a polariton BEC has not yet been explored. In this work, we are interested in accessing the dynamics of the second-order time correlation function during the spontaneous BEC formation under nonresonant excitation in a CdTe microcavity.

\begin{figure}
  \includegraphics[width=0.48\textwidth]{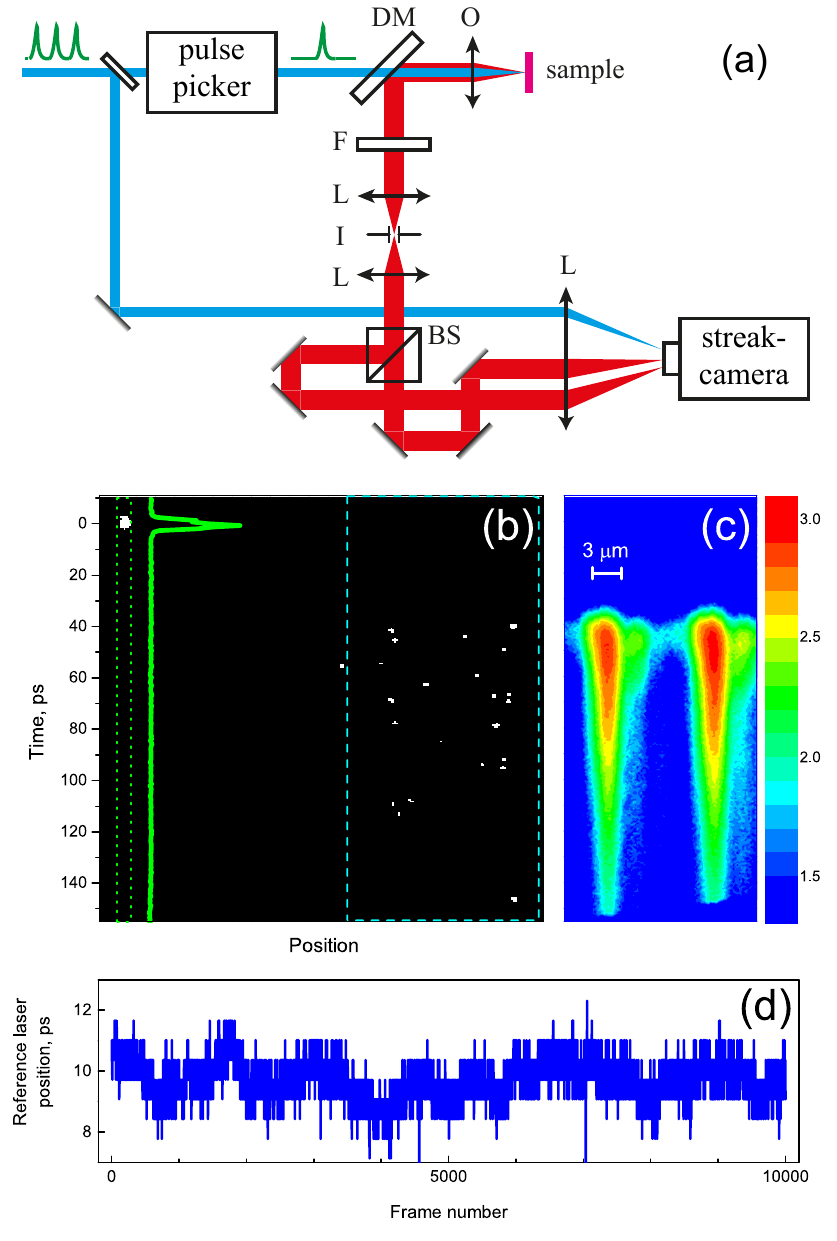} 
  \caption{(a) Sketch of the experimental setup. DM - dichroic mirror, O - objective, F - bandpass filter, L - lens, I - iris diaphragm, BS - beamsplitter. (b) Single photon data acquired during a single shot of the streak-camera. White pixels correspond to intensity above photon counting threshold. A spot in the upper-left corner is a reference laser pulse. Green reference laser temporal profile was calculated from the data in green dotted rectangle. (c) Photon counts in the cyan dashed rectangle in (b) integrated over 100'000 single shots and plotted in logarithmic scale. Real-space resolution is given on the plot. (d) Long-term time shift determined from the reference laser position over the first 10'000 frames.}
  \label{fig:fig1}
\end{figure}

We report high temporal resolution $g^{(2)}$ measurements of the formation of a polariton BEC from a thermal polariton population. These results rely on a C5680 streak-camera with an S25 photocathode. The sample is the CdTe planar microcavity which was used in our previous works\cite{Kasprzak_Nature2006,Kasprzak_PRL2008}. It contains 16 quantum wells and shows a Rabi splitting of 26 meV. The sample was held in a helium-flow optical cryostat at 4.2 K and was excited through a 0.5 numerical aperture microscopic objective with non-resonant linearly polarized pulses from a Ti:Sapphire laser with central wavelength 697 nm, a pulse length around 300 fs and a top-hat beam profile. The excitation power was 10 times higher than the threshold power required for BEC formation ($P_{thr}$). Here, we studied a single BEC with a size of 15 $\mu$m in a potential landscape with an energy minimum in its center. The sample photoluminescence (PL) at 740 nm was collected in reflection geometry by the same microscope objective and a dichroic mirror was used to separate the PL from the excitation beam (Fig. \ref{fig:fig1} (a)). To further cut off the laser light scattered by the sample, a bandpass filter was used.

For measuring $g^{(2)}$, the emission of the sample was sent to a beamsplitter, realizing a HBT detection scheme, and then to the streak-camera entrance slit (Fig. \ref{fig:fig1} (a)). To measure the second-order time correlations, one needs to study photon statistics after each single excitation event of the sample. Previously, a slow horizontal sweep of a streak-camera was used to separate sample emission events coming from different pulses \cite{Wiersig_Nat2009}. In contrast, for each frame recorded by the streak-camera, we only send on its slit the photons emitted after a single excitation pulse. For this, we select single laser excitation pulses using a pulse picker triggered by the streak-camera electronics. This ensures us that during one streak-camera frame exposure only one laser pulse hits the sample, and hence only one BEC realization is probed. The pulse picking system used was based on a Conoptics electrooptical modulator. An example of a single event recorded by the streak-camera is shown in Fig. \ref{fig:fig1} (b). After recording each photon event for a given frame, and integrating over 100'000 single pulses, the emission shown in Fig. \ref{fig:fig1} (c) clearly reproduces the dynamics of a polariton BEC.

The calculation of $g^{(2)}$ in this configuration is done in the standard way:
\begin{equation*}
	g^{(2)}(t_{1},t_{2}) = \frac{\left< n_{1}(t_{1})n_{2}(t_{2}) \right>}{\left< n_{1}(t_{1})\right> \left< n_{2}(t_{2}) \right>}
\end{equation*}

where $n_{1,2}$ correspond to the counts in the left (right) arm of the HBT experiment at times $t_{1,2}$ spanning the range of the detector. The value $g^{(2)}(\tau)$ is defined for $\tau = t_{1}-t_{2}$ and $g^{(2)}(0)$ occurs when $t_{1}=t_{2}$. The error is calculated from the standard error of the local intensities and of the number of coincidences ($\Delta\left< n_{1}\right>$, $\Delta\left< n_{2} \right>$ and $\Delta\left< n_{1} n_{2} \right>$). The use of HBT setup prevents possible problems with the counting of photons with short time delay intervals \cite{Schmutzler_PRB2014}.

To obtain small enough experimental error, we have to accumulate statistics over at least one hundred thousand frames. This requires a long experimental time during which locking between the streak-camera sweeping pulses and the laser pulses can fluctuate. This leads to shifts of the time axis of the streak-camera, which may cause a significant reduction of the temporal resolution. To overcome this, we sent a tiny fraction of the exciting laser pulse on the streak camera (Fig. \ref{fig:fig1} (a)) in order to determine precisely the fluctuations of the laser position. Though the streak-camera was operating in a photon counting regime, the relatively strong emission of the reference laser produced a bunch of counts, forming a clear spot on the streak-camera screen (like in the upper-left corner of Fig. \ref{fig:fig1} (b)). The center of this spot was computed for each frame, allowing us to quantitatively assess the time shift with very high accuracy (Fig. \ref{fig:fig1} (d)) and correct single photon counts, shifting them along the time axis. It is clearly seen from Fig. \ref{fig:fig1} (d), that the measured time shift exhibits oscillations. Although it could have been described as an extra Gaussian jitter, our simple and efficient correction routine allowed us to decrease very significantly the effects of this timing jitter on our experiments.

\begin{figure}
  \includegraphics[width=0.48\textwidth]{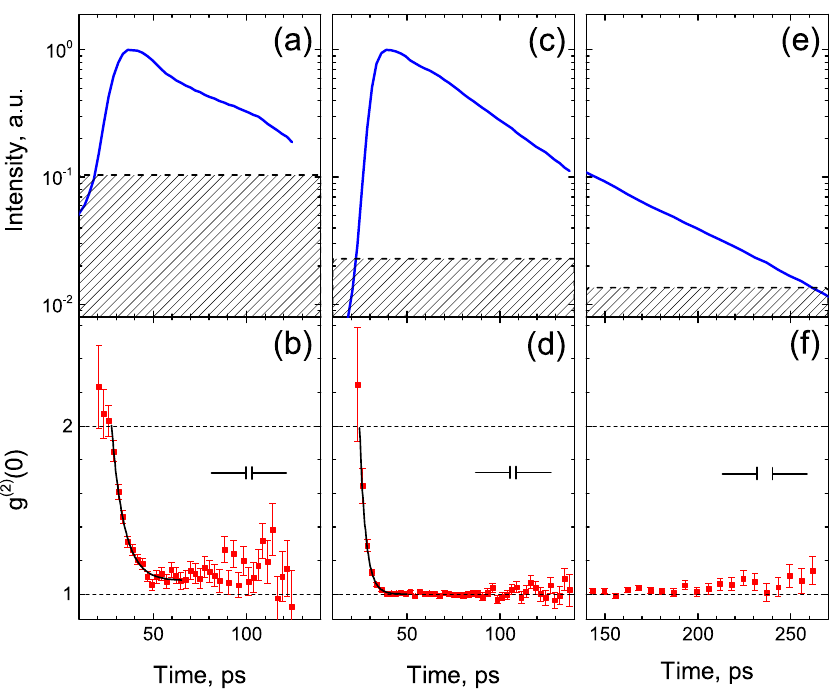} 
  \caption{(a) Dynamics of PL intensity and (b) $g^{(2)}(0)$ for a single BEC without real-space filtering. (c-d) The same, but with real-space filtering. (e-f) The same as (c-d), but recorded using slower sweep range of the streak-camera (only the tail of dynamics is shown). The temporal resolution is given on the plots and equals 2.2 ps for (a-d) and 10.6 ps for (e-f). Single photons integrated in time bins of 2.6 ps for (b,d) and 6.2 ps for (f). Shaded areas indicate regions where signal-to-noise ratio is below 20/1, for which significant distortion of measured $g^{(2)}(0)$ is expected. Black thick lines in (b) and (d) are exponential fits of $g^{(2)}(0)$ with decay time of 6.7 ps for (b) and 3.2 ps for (d).}
  \label{fig:fig2}
\end{figure}

Another source that causes deviations of measured $g^{(2)}(0)$ values from the real ones originates from the dark counts. Their presence can be described as an additional uncorrelated mode \cite{Assmann_OE2010} and is known to pull the real $g^{(2)}(0)$ value towards 1. The role of dark counts depends on signal-to-noise ratio (SNR) \cite{Assmann_PRB2010} and starts to be significant at SNR below 20/1. To avoid misleading data, we do not present here any results of $g^{(2)}$ measurements for SNR below 20/1 (see the shaded areas in Fig. \ref{fig:fig2}).


The dynamics of the normalized average PL intensity and the calculated $g^{(2)}(0)$ values for the entire BEC are presented in Fig. \ref{fig:fig2} (a,b). When the build-up of the BEC starts, we see a clear transition of $g^{(2)}(0)$ from a value around 2 to roughly 1.1 (Fig. \ref{fig:fig2} (b)), confirming the transition from a thermal population of polaritons to a coherent state. While the value of $g^{(2)}(0)$ approaches 1, it does not reach 1. The same behavior was previously reported for HBT measurements using avalanche photodiodes \cite{Kasprzak_PRL2008,Love_PRL2008} and has been described as a consequence of the interactions of the condensate with the thermal reservoir \cite{Schwendimann_PRB2008}. However, we have found that, when we use real-space filtering (see Fig. \ref{fig:fig1} (a)) to observe only the PL from the central part of BEC (9 $\mu$m in diameter), we observe a drastic improvement in the coherence statistics (Fig. \ref{fig:fig2} (d)) giving $g^{(2)}(0) = 1.005 \pm 0.002$. Such high degree of coherence persists for at least 100 ps (Fig. \ref{fig:fig2} (d)). To further trace the dynamics of $g^{(2)}(0)$, we use a slower sweep of the streak-camera (Fig. \ref{fig:fig2} (e-f)). In this mode, when the intensity of the sample emission reaches its maximum, we have too many photons coming onto the streak-camera screen, and the photon counting routine fails to count them. This is the reason why we show only the tail of the decay in Fig. \ref{fig:fig2} (e-f). As can be seen from Fig. \ref{fig:fig2} (f), $g^{(2)}(0)$ stays at a value close to 1 until the end of the experiment.

At the very end of the experiment (Fig. \ref{fig:fig2} (f)) we observe slight increase of the value of $g^{(2)}(0)$. To explain this, we recall that the polariton condensate lifetime is not given by the lifetime of the polaritons themselves (a few ps only) but by the feeding from the excitonic reservoir \cite{Lagoudakis_PRL2010, Manni_PRL2011vort}, which has been observed to last for more than 100 ps. When the population of the exciton reservoir gets too low, the feeding of the BEC is no longer efficient, which leads to a breakdown of the condensate. At this point, emission starts to be dominated again by the noncondensed thermal polaritons and this should give a value of $g^{(2)}(0) = 2$. However, it is hardly possible to observe this transition because the breakdown of the BEC should occur at a number of polaritons per state of the order of 1. In this regime, SNR will be too low to measure $g^{(2)}(0)$ with any reasonable precision. Though we cannot see how the emission comes back to thermal statistics, we presumably can observe beginning of this process during the last 30 ps on the experiment as shown in Fig. \ref{fig:fig2} (f).

The observed difference between $g^{(2)}(0)$ measurements with and without real-space filtering of the BEC image cannot be ascribed to the thermal population of polaritons because all the measurements were performed at excitation well above threshold ($10 P_{thr}$) and the fraction of noncondensed polaritons cannot reach any significant value. Instead, we relate the measured $g^{(2)}(0)$ values to the first order spatial correlation function $g^{(1)}$ of the polariton condensate. This function describes the degree of spatial coherence over the size of BEC and is known to decrease from 1 in the center of the two-dimensional polariton condensate to 0 at its edges \cite{Krizhanovskii_PRB2009}. We suppose that the loss of coherence between different parts of a large BEC is responsible for a deviation from $g^{(2)}(0) = 1$ (Fig. \ref{fig:fig2} (b)). Real-space filtering of PL allows to observe emission only from the center of BEC, where the condensation occurs first due to the minimum of potential landscape. Hence, the central region is characterized by higher spatial coherence, which leads to the observation of a value of $g^{(2)}(0)$ almost equal to 1, namely $1.005 \pm 0.002$. The small difference from 1 can be caused by both the thermal fraction of polaritons and a $g^{(1)} < 1$ across the central region of the condensate.

\begin{figure}
  \includegraphics[width=0.48\textwidth]{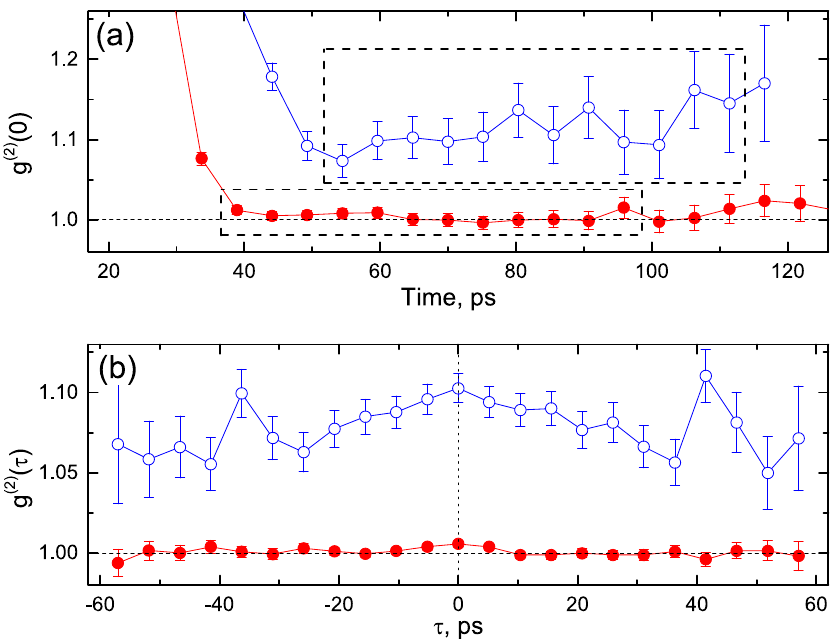} 
  \caption{(a) Dynamics of $g^{(2)}(0)$ for a single BEC with (red filled points) and without (blue open points) real-space filtering. (b) Results of $g^{(2)}(\tau)$ measurements for data within dashed rectangles in (a). Single photons integrated in time bins of
5.2 ps. Temporal resolution is the same as in Fig. \ref{fig:fig2}}
  \label{fig:fig3}
\end{figure}

As evidenced in Fig. \ref{fig:fig2} (d), polaritons form the condensed state with a characteristic time of 3.2 ps. This value is consistent with time-resolved real-space interferometry measurements \cite{Manni_NComm2012}, however, being close to the time resolution of the streak-camera (2.2 ps), it should be considered as an upper limit. This transition time is defined not only by the process of condensation across the region observed, but also by the time that is required to synchronize the phase across the whole BEC. The latter is governed by Kibble-Zurek-like mechanisms \cite{Kibble_JPA1976,Zurek_PRep1996,Manni_NComm2013} and is based on the propagation of the phase across the condensate with a speed approximately equal to the speed of sound, which is of order of a few $\mu$m/ps \cite{Nardin_PRL2009}. This implies that synchronization of the phase across a large BEC \cite{Lagoudakis_PRL2008} should take more time. This fact is confirmed by the data on Fig. \ref{fig:fig2} (b). On this plot, we see that the stabilization of the phase of the full BEC takes $6.7 \pm 0.6$ ps, longer that time required to stabilize the phase around the central region (Fig. \ref{fig:fig2} (d)), which confirms our hypothesis.

In Fig. \ref{fig:fig3} (a) we have a closer look on the dynamics of $g^{(2)}(0)$ for both the entire condensate and for its central region only. Results for $g^{(2)}(\tau)$, calculated from the weighted mean of $g^{(2)}(t_{1},t_{2})$ for the BEC emission during the time region marked by the dashed rectangle on Fig. \ref{fig:fig3} (a), are shown in Fig. \ref{fig:fig3} (b). A clear reduction in $g^{(2)}(\tau)$ is observed for the PL from only the central region of BEC. $g^{(2)}(\tau)$ is supposed to reach value of 1 for $\tau$ significantly larger than the coherence time. This transition cannot be seen for spatially filtered PL since even at $\tau=0$ $g^{(2)}(\tau)$ is already very close to 1. As for the PL without real-space filtering, we do not observe $g^{(2)}(\tau)$ getting to 1, but can notice a clear reduction of $g^{(2)}(\tau)$ for long delays. This gives us a lower estimate of the BEC coherence time of the order of 60 ps (the time limit of the experiment). This timescale is similar to the one given by the results of time-resolved measurements of $g^{(1)}$ on the same sample \cite{Love_PRL2008}. Such long coherence time is presumably defined by the lifetime of an excitonic reservoir that repopulates the BEC, which is of order of few hundreds of picoseconds \cite{Lagoudakis_PRL2010, Manni_PRL2011vort}.

In summary, we have measured the ultrafast dynamics of second-order time correlation functions $g^{(2)}(0)$ and $g^{(2)}(\tau)$ in a spontaneously formed BEC of exciton-polaritons. The process of BEC formation from a thermal non-condensed state with $g^{(2)}(0) = 2$ to a polariton BEC with $g^{(2)}(0) = 1$ is clearly observed. The transition occurs within less than 3.2 ps. Once condensed, polaritons show $g^{(2)}(0)$ persisting around 1.0 with high precision for at least several hundred picoseconds. Finally, we evidence, contrary to previous studies, that there are no significant thermalizing interactions between the reservoir and the polariton BEC, which is confirmed by the measured value of $g^{(2)}(0) = 1.005 \pm 0.002$. Accessing coherence properties of polariton BECs by dynamically tracking $g^{(2)}(0)$ might reveal a generic approach for study of ultrafast processes in polaritonic systems, like Josephson \cite{Lagoudakis_PRL2010} or Rabi \cite{Norris_PRB1994} oscillations, as well as quantum processes like intensity squeezing \cite{Boulier_NComm2014} or polariton blockade \cite{Liew_PRL2010}.

\begin{acknowledgments}
We would like to thank Hugo Flayac and Naotomo Takemura for fruitful discussions. The work is supported by the Swiss National Science Foundation under project No. 153620, the Quantum Photonics National Center of Competence in Research No. 115509 and from European Research Council under Polaritonics project No. 219120. The POLATOM network is also acknowledged.
\end{acknowledgments}

\bibliography{bibfile}

\end{document}